\documentclass[10pt]{iopart}
\usepackage{epsf,citesort}
\usepackage{graphicx,amssymb}

\begin{document}
\title{Simulating Colloid Hydrodynamics with Lattice Boltzmann}
\author{M. E. Cates$^1$, K. Stratford$^2$, R. Adhikari$^1$, P. Stansell$^1$, J.-C. Desplat$^2$, I. Pagonabarraga$^3$, A. J. Wagner$^4$}
\address{$^1$ School of Physics, JCMB Kings Buildings, The University of
Edinburgh, Mayfield Road, Edinburgh EH9 3JZ, United Kingdom}
\address{$^2$ EPCC, JCMB Kings Buildings, The University of
Edinburgh, Mayfield Road, Edinburgh EH9 3JZ, United Kingdom}
\address{$^3$ Departament de Fisica Fonamental, Universitat de Barcelona,
Av. Diagonal 647, 08028 Barcelona, Spain}
\address{$^4$ Department of Physics, North Dakota State University, ND58105 USA}

\newcommand{\x}{{\bf x}}
\renewcommand{\c}{{\bf c}}
\renewcommand{\v}{{\bf v}}
\renewcommand\Re{\mbox{\rm Re}}
\newcommand\Pe{\mbox{\rm Pe}}
\newcommand\Ca{\mbox{\rm Ca}}

\begin{abstract}
We present a progress report on our work on lattice Boltzmann methods for colloidal suspensions. We focus on the treatment of colloidal particles in binary solvents and on the inclusion of thermal noise. For a benchmark problem of colloids sedimenting and becoming trapped by capillary forces at a horizontal interface between two fluids, we discuss the criteria for parameter selection, and address the inevitable compromise between computational resources and simulation accuracy.
\end{abstract}

\maketitle

\section{Introduction}
The lattice Boltzmann equation (LB equation, or LBE) is a widely used lattice formulation of
fluid mechanics \cite{succi}. It offers a faithful discretisation of
the Navier Stokes equation of isothermal, near-incompressible fluid flow, and is very well adapted to parallel
computation \cite{parallel}. Although used for large-scale fluid
dynamics simulations such as flows around aircraft wings \cite{aircraft}, the
LBE approach is particularly adapted to simulating mesoscopic problems
\cite{discreteproc}. These include, for example, porous medium flows, and flows of
complex and multicomponent fluids with microstructure
\cite{warren,kendon,njp,coveney,ladd}. The latter can be modelled using various
extensions of the basic algorithm for a single component fluid \cite{swift,coveneyalg,ladd}. 

In this paper we outline some recent progress towards the creation and use of a versatile LB code for colloids in single-phase and (particularly) biphasic
fluid solvents \cite{stratford}. The LB algorithm is not wholly intuitive, and the mapping of simulation parameters onto real ones
has to be carried out with some care, with attention paid to possible sources of systematic error. For problems involving coarsening of binary fluids, the required validation is provided in \cite{kendon}; below we discuss some of the additional considerations that arise for colloids (Section \ref{parameters}).
In addition, because of the mesoscopic length scales involved, such a code must allow a proper treatment of thermal noise. We have recently identified and resolved a longstanding difficulty with the incorporation of noise in LB \cite{adhikari}, and this is discussed in Section \ref{ghosts}. Before turning to these topics we briefly review, in Sections \ref{onephase}--\ref{collbinary}, the LB algorithm for a single phase fluid, for binary fluids, and for colloids. 

\section{The LB Algorithm for a single phase fluid}
\label{onephase}
LB works one level beneath the usual equations of hydrodynamics, at the level of
the collisional and propagating dynamics of distribution functions. The distribution function $f_i(\x,t)$ at lattice site $\x$ can be thought of as the density of (fictitious) fluid particles with velocity $\c_i$ at this site. The velocity set is chosen so that in one timestep $\Delta t$, the displacement $\c_i \Delta t$ represents either the displacement to a neighbouring lattice site or the null displacement. We use a cubic D3Q15 lattice meaning a three dimensional lattice with 15 velocities at each site
\cite{ludwig}. 

Note that a larger number of velocities, and hence of distribution functions, are introduced at each lattice site than is strictly needed to create the correct number of hydrodynamic degrees of freedom for fluid-mechanical purposes.
These extra degrees of freedom are required so as to ensure rotational and Galilean invariance \cite{succi}; they lead to `ghost modes' which feature strongly in Section \ref{ghosts} below. Setting these ghost modes aside, the hydrodynamic fields (which are all that matter for athermal fluid mechanics) are represented by various moments of the distribution functions. For example the local density of the fluid is the zeroth moment
\begin{equation}
\rho(\x,t) = \sum_if_i(\x,t) 
\label{density}
\end{equation}
whereas the local momentum density ${\bf g} = \rho\v$ is given by
\begin{equation}
\rho\v = \sum_i f_i(\x,t)\c_i
\label{momentum}
\end{equation}
and the momentum flux (or kinetic stress tensor) is given by
\begin{equation}
{\mbox{\boldmath $\Pi$}} = \sum_if_i \c_{i}\c_{i}
\label{momentumflux}
\end{equation}
where $\c_i\c_i$ is a dyadic product. 

Although LB actually describes a compressible fluid, in using it one always ensures that the Mach number is kept small so that the flow is nearly incompressible. 
We choose our unit of mass so that $\rho = 1$ for a quiescent fluid, and choose the lattice parameter as the unit of length, and the timestep as the unit of time ($\Delta t = 1$). This defines a set of lattice units (LU). The interconversion between these and real physical units raises interesting issues in parameter selection (see Section \ref{parameters}). 

The LB algorithm updates the distribution functions $f_i$ by a combination of
`streaming' and `collision'. These are often thought of as two different steps
in the algorithm. Streaming passes each distribution function $f_i$ to the neighbouring site appropriate to the velocity $\c_i$ that it governs; whereas
collision is an on-site update of the various $f_i$ which conserves mass and momentum (but not energy), creates dissipation, and allows diffusion of momentum. (The diffusivity of momentum is the kinematic viscosity $\eta/\rho$ of the fluid.) In combination, the streaming and collision steps may be written as
\begin{equation}
f_i(\x+\c_i,t+1) - f_i(\x,t) = \sum_j{L}_{ij}\left(f_j(\x,t) -f_j^0(\x,t)\right)
\label{LBE}
\end{equation}
where $L_{ij}$ is a collision matrix, often chosen as the lattice BGK matrix
$\delta_{ij}/\tau$ with $\tau$ a relaxation time. In Eq.\ref{LBE}, $f_j^0(\x,t)$ is an equilibrium distribution \cite{succi} which itself depends on the local values of
$\rho(\x,t)$ and $\v(\x,t)$, so that the collision process
(even with the stated BGK form for $L_{ij}$) is not diagonal among the $f$'s. (The nontrivial structure of its eigenmodes will be important in Section \ref{ghosts}.)

Meanwhile, the fluid shear viscosity is given (in LU) by 
\begin{equation}
\eta = c_s^2(2\tau-1)\rho/2
\label{viscosity}
\end{equation}
where $c_s$ is the sound speed; for the D3Q15 lattice that we use, $c_s = 1/\sqrt{3}$ in LU. (There is also an unimportant bulk viscosity, which for this lattice is $\zeta  = 2\eta/3$.)
In practice $\tau = 1$ is numerically efficient: in this case the collision resets the $f_i$ to local equilibrium, each time step. Larger values of $\tau$ are possible, but can bring numerical problems. In particular it is not possible for the momentum to be transported across a given
distance by diffusion any faster than it can get there by sound modes (at speed $c_s$), so that $\eta/\rho$ values large compared to $c_s$ will give incorrect momentum transport at short distances (and/or unwanted non-Newtonian effects \cite{wagnervisco}). 
Viscosity values much smaller than unity can be used, and these are very helpful in studying phase separation of binary fluids at high Reynolds number \cite{kendon,njp}. 

The basic LB algorithm, as just described, is sufficent to model isothermal flow of a single-phase fluid with a variety of boundary conditions \cite{succi}. In the continuum limit for the bulk fluid, one recovers (with Greek suffices for
cartesian components)
\begin{eqnarray}
\label{MASSCONSERVATION}
\partial_t \rho + \nabla_{\alpha}g_{\alpha}=0\\
\label{MOMENTUMCONSERVATION}
\partial_t g_{\alpha}+\nabla_{\beta}\Pi_{\alpha\beta}=0\\
\label{CONSTITUTIVE}
\Pi_{\alpha\beta}=g_{\alpha}v_{\beta}+p\delta_{\alpha\beta}-\eta_{\alpha\beta\gamma\epsilon}\nabla_{\gamma}v_{\epsilon}\end{eqnarray}
Here $v_{\alpha}=g_{\alpha}/\rho$ is the local fluid velocity, $p$
is the pressure in a quiescent fluid (given in LB by an ideal gas equation of
state $p=\rho c_s^2$), and
$\eta_{\alpha\beta\gamma\epsilon} = \eta(\delta_{\alpha\gamma}\delta_{\beta\epsilon} + \delta_{\beta\gamma}\delta_{\alpha\epsilon} -\frac{2}{d}\delta_{\alpha\beta}\delta_{\gamma\epsilon}) + \zeta \delta_{\alpha\beta}\delta_{\gamma\epsilon}$ 
is the tensor of viscosities appropriate to an isotropic Newtonian fluid.
It is possible to impose shearing boundary conditions through an imposed fluid velocity or stress at a pair of walls parallel to lattice directions; less obviously (given the existence of the underlying lattice) one can also introduce Lees-Edwards-type sheared periodic
boundary conditions \cite{wagnerlees}. One limitation is that the fluid velocity throughout the system must remain small compared to the sound speed. This is a stronger condition than that of small Mach number (small compressibility) because it implies an absolute reference frame against which the flow speed is limited. In shear flow, however, this limit can be overcome by judicious use of multiple Lees-Edwards planes \cite{wagnerlees}.

\section{Binary fluids in LB}
Binary fluids can be handled by an extension to the above approach \cite{swift,ludwig}, in which a second set of distribution functions $g_i(\x,t)$ is introduced. Analagously with Eqs.\ref{density},\ref{momentum},{\ref{momentumflux}, the low-order moments govern the compositional order parameter $\phi(\x,t)$ and its advective and diffusive
fluxes. The equilibrium distribution $g_i^0$ involves an order parameter mobility and a chemical potential $\mu$ which derives from a well-chosen free energy functional $F(\phi,\nabla\phi)$, which we take to be of Landau-Ginzburg form \cite{kendon,ludwig}. A second relaxation time is also introduced although, rather unintuitively, the order parameter mobility, so long as it is independent of $\phi$, can be varied without changing this relaxation time (varying it instead by the choice of $g_i^0$). In our codes this second relaxation time is set to unity to optimise numerical efficiency; thus the $g_i$ are reset to local equilibrium
every time step. 

The binary fluid LB algorithm is in many ways less satisfactory than that for a single fluid. For example if a static droplet of one fluid is surrounded by another, there are weak violations of detailed balance due to `spurious fluxes' which arise from the fact that the proper conservation laws for the order parameter are not built in at a deep enough level. In future it may be possible to improve this \cite{adhikari2}, but meanwhile the algorithm gives useful results for phase separation dynamics \cite{kendon,crossover,njp,wagnerdrops}, phase separation with a temperature ramp \cite{wagnerramp}, droplet breakup under shear \cite{wagnerlynn,droplets} and related problems. In combination with Lees-Edwards boundary conditions it allows the problem of sheared spinodal phase separation to be studied; we hope to publish more on this soon \cite{stansell}. 

One advantage of LB over other methods for binary fluids is that, by working at the level of distribution functions, it avoids the hydrodynamic singularities that arise during pinchoff and topological reorganisation of fluid domains \cite{eggers}. Such singularities are smoothed out by order parameter diffusion; obviously this is only helpful under conditions where the singularities are a nuisance, rather than the controlling physical effect. The binary fluid LB does require careful parameter steering to avoid spurious effects such as anisotropy of the fluid interface, and excessive order-parameter diffusion in regions where this should be negligible compared to advection \cite{kendon}.

\section{Colloid hydrodynamics in LB}
The introduction of moving solid objects in LB is a somewhat complicated procedure \cite{ladd,aidun,heemels,nguyenladd}. Firstly the objects have to be mapped onto the lattice grid; but since they may be large and slowly moving, their positions cannot be moved by discrete jumps of one lattice unit. Hence the colloids are modelled off lattice, but each defines a set of links of the lattice crossed by the surface of the colloidal particle, and this link-set is subject to discrete modification at each time step. The fluid distribution functions at these link nodes is handled via a `bounce-back' procedure, in which velocities that would correspond to crossing into the colloid are reflected back into the fluid according to certain rules. The resulting force and torque on the colloid is found, and used to update its velocity and angular velocity. This information is not only used to update the colloid positions each time step, but fed back into the bounce-back prescription for the next step (which depends on the local velocity of the moving solid boundary).

In our work we adopt and modify the prescription of \cite{nguyenladd}. In contrast to earlier work \cite{ladd} where the interior of each colloid was filled with a fictitious fluid, here the bounce-back rules treat the colloidal particles as truly solid. This distinction is important when dealing with binary solvents, in which one needs to develop rules for the compositional order parameter $\phi$ as well as the other hydrodynamic fields \cite{stratford}. Due to the changing discretisation of the boundary links, the shape of each colloid is in effect changing slightly as it moves across the lattice; but it is possible to calibrate this effect and get acceptable hydrodynamic behaviour for surprisingly small colloids, for example of radius $a  = 2.4$ lattice units. However, the discrepancy between the real particle radius and the measured hydrodynamic one increases with viscosity \cite{stratford}, for reasons that are currently under investigation. 

An alternative route would be to treat colloids as point particles and couple each of these locally to the fluid velocity, with a friction constant fixed by Stokes law. This is 
a very useful approach to polymer hydrodynamics, where the procedure is applied to each bead in a long chain \cite{dunweg}. Since that problem is known to be dominated by far-field effects (leading to Zimm dynamics rather than Rouse), the fact that the near-field flow is treated inaccurately by such an approach does not matter. However, for colloid hydrodynamics it certainly does matter -- for example, this method would not give accurate results for sedimentation of a small group of particles at separations comparable to their diameters, even if no very close contacts (lubrication forces) came into play. Hence for a general-purpose colloidal hydrodynamics code there is no way to bypass the bounceback procedure, or some equivalently elaborate scheme, in which forces are explicitly distributed over the surface of each colloid. 

An important part of the hydrodynamics of interacting colloids involves near-field lubrication forces. At small surface-to-surface distances these have a strong divergence in the normal component (and a much weaker one in the tangential) which LB can only resolve down to separations between colloidal surfaces of order one lattice spacing. These strong local contributions, unlike the other hydrodynamic terms, are pairwise additive across the colloid-colloid contacts. Hence their absence from LB can be rectified by patching in, at short distances $r<r_c$ only, a velocity-dependent lubrication force \cite{nguyenladd}, acting directly between the particles. (In our code, we so far treat the normal component only \cite{stratford}.) However this brings its own difficulties; because the force is velocity dependent, an implicit update scheme is now essential for the colloid dynamics. The computational time required for this scales badly with the size of any `hydrodynamic clusters' (clusters of colloids mutually linked together by separations less than $r_c$) and such clusters can get very large at high concentrations. 

In effect, in this regime, one is inverting the pairwise-additive colloidal drag matrix to get a mobility matrix, every timestep. Indeed algorithms for colloid hydrodynamics exist that do only this, within a pairwise lubrication approximation \cite{melrose}, ignoring the far-field contributions that in our case are handled by LB. (This far field can be handled within Stokesian dynamics \cite{brady} which, like the pairwise lubrication algorithm but unlike LB, assumes creeping flow.) Such work shows that deviations from the hard-sphere potential are critical in determining, for example, jamming behaviour under shear \cite{melrose}. 

In principle an LB algorithm with lubrication corrections should give results that can encompass this strong clustering limit without sacrificing the long-distance aspects of many-body hydrodynamics \cite{nguyenladd}. However, in view of the computational scaling issue, we have not attempted to explore this aspect, and prefer instead to bypass the lubrication problem entirely. This can be done by introducing a strong, short-range repulsion between colloids designed to ensure that interparticle separations $r<r_c$ are rare, so that the bad scaling does not arise. Of course such a force is often actually present, for example, in colloids interacting by a screened Coulomb force. We expect this additional repulsion to be unimportant for some scientific issues (including those we address below) but not others (e.g. not hydrodynamic jamming under shear \cite{melrose}).

\section{Colloids in Binary Solvents}
\label{collbinary}

A colloid in a pair of solvents is said to exhibit neutral wetting when the solid-fluid interfacial tension $\tilde\sigma$ is the same for both solvents. This corresponds to a contact angle of $\pi/2$. Such colloids are strongly surface active; the interfacial energy is $4\pi a^2\tilde\sigma$ for a colloid wherever it is placed in the two fluids, but placing it symmetrically across the interface between them reduces the area of fluid-fluid contact by $\pi a^2$. Hence a neutrally wet colloid is bound to the fluid-fluid interface by an energy $\pi a^2\sigma$ with $\sigma$ the fluid-fluid interfacial tension. In most circumstances this quantity vastly exceeds $k_BT$ and therefore colloidal adsorption to the interface is effectively irreversible. This is the basis of several technologies involving emulsions stabilised by solids, which are generally called `Pickering emulsions' \cite{aveyard}. Studying new variants of  these is a primary motivation for our development of an LB code to handle colloids in binary solvents. Our current implementation of the code is restricted to neutral wetting, but similar physics should be seen for a range of angles around $\pi/2$. (The issue of how to vary the contact angle away from neutrality is understood in principle, and implemented but not yet tested in our colloid codes \cite{ludwig,stratford}.)

\section{Benchmark Problem and Parameter Selection}
\label{parameters}
Before using a somewhat complicated algorithm (such as LB) to gain quantitative results for an equally complicated problem (such as colloids in binary solvents) it is important to complete a range of benchmark tests so as to give qualitative and quantitative insights into sources of systematic error. One must also develop a strategy for choosing simulation parameters to get close enough (given these sources of error) to some experimentally realisable system of interest. For binary fluids undergoing coarsening, extensive parameter testing is reported by Kendon et al \cite{kendon}, and we build on this wherever we can. But as far as colloids in binary solvents are concerned, the task is not yet finished: the results presented below are preliminary only.

To illustrate the principles involved, let us consider a simple geometry in which a suspension of colloids, at low volume fraction, sediments under gravity within a stratified pair of layers of two immiscible fluids of equal viscosity. Figure \ref{bench} gives a series of snapshots of this process; each shows the bottom half of a system with periodic boundary conditions. Thus, one fluid lies below the interface, another on top (and out of view above this is a second interface which restores periodicity in the vertical direction). The colloidal particles are initially placed at random, so that some of them lie across the interface but most do not. They then fall under gravity and, for the parameters selected here, they all end up attached to the interface.

\begin{figure}
\begin{center}
\includegraphics[width=12cm]{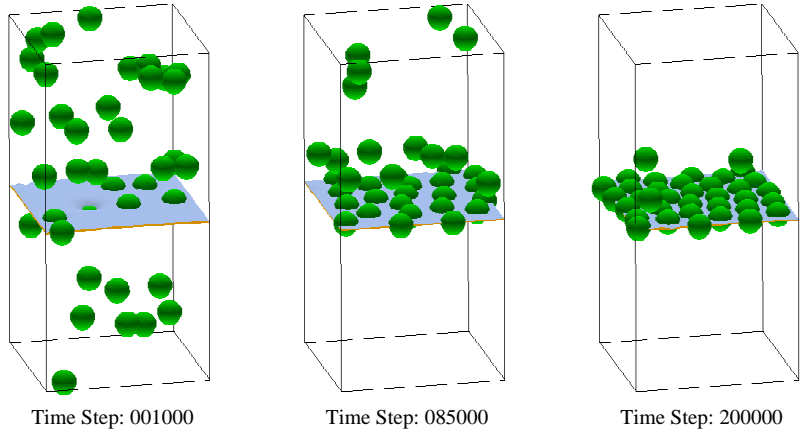}
\end{center}
\caption{\label{bench}Snapshot configurations for the benchmark sedimentation problem with Ca = 280 and Re = 0.02.} 
\end{figure}

As is proper in a fluid mechanics problem, we address the physics of the situation by identifying some relevant dimensionless numbers. Indeed, these are particularly helpful in mapping between lattice units and the real world.
The relevant numbers include the gravitational Reynolds and Peclet numbers \cite{chaikin}:
\begin{equation}
\Re = \frac{v_{sed} a \rho}{\eta}
\end{equation}
where $v_{sed} = \Delta mg/6\pi\eta a$ is the sedimentation velocity, and
\begin{equation}
\Pe = \frac{v_{sed} a}{D}
\end{equation}
where $D = k_BT/6\pi\eta a$ is the diffusion constant of a colloid.
Note that for terrestrial gravity $g$ and typical density mismatch ($\Delta mg/mg \sim 0.1 - 1$, with $m = \rho 4\pi a^3/3$) one has Re typically of order $10^{-7} (a/a_\mu)^3$ and Pe of order $(a/a_\mu)^4$. 
Here we have introduced a conventient length scale $a_\mu \equiv 1\mu$m.
Thus a colloid of, say, $a = 10\mu$m has negligibly
small inertial effects (small Re) and also negligibly small diffusion (large Pe), when falling under gravity. For $a = 100$nm, Re is still negligible but Pe is now small. Note that Pe may also be written as $a/h$, where $h = k_BT/\Delta m g$ is the gravitational decay length for the colloidal concentration in barometric equilibrium.

The fluid-fluid interfacial tension $\sigma$ introduces another dimensionless parameter, which we can write as a (gravitational) capillary number
\begin{equation}
\Ca = \frac{v_{0}}{v_{sed}} 
\end{equation}
Here, $v_0$ is an intrinsic velocity scale $\sigma/\eta$ which governs the dynamics of a disturbed interface between fluids \cite{kendon}. Thus for large Ca, the interface can adapt easily to the arrival of particles but for small values the particles are likely to break through it as they sediment. Indeed, Ca $= 6\pi \sigma a/\Delta mg$, so that, to within an order-unity factor, Ca also determines whether the force of attachment of a particle to the fluid-fluid interface is enough to hold it there against gravity. (At large Ca sedimenting particles should gather at the interface whereas for small Ca they will fall off it even if placed there gently.) Note that for typical binary fluid parameters, Ca $\simeq 3\times 10^{8}(a/a_\mu)^{-2}$.
This large value (at $a\simeq a_\mu$) reflects the strength of the binding to fluid-fluid interfaces: particle sizes of order millimetres are needed before gravity will detach a neutrally wetting particle from, say, an oil-water interface. Note that a similar dimensionless quantity arises for the gravitational equilibrium of a pendant droplet of one fluid suspended against gravity in another by capillary forces; in this context, Ca$^{-1}$ is usually
called the Bond number.

Under conditions where sedimentation is strong and diffusion weak, we can set Pe to be infinite. This is in fact the default position in LB unless noise is explicitly added. (A way to do that is described in Section \ref{ghosts}.) More interesting is the role of the Reynolds and capillary numbers; we address these in turn. 

As outlined above, for typical colloid parameters Re is extremely small. However, LB works by solving dynamically a discretisation of the full hydrodynamic equations 
\ref{MASSCONSERVATION},\ref{MOMENTUMCONSERVATION},\ref{CONSTITUTIVE}.
This means in practice that Re can never be made fully realistic for colloids, as we now explain. A reasonable duration of a simulation is of order $10^4$ (or perhaps $10^5$) timesteps; a reasonable lattice size of order $128^3$
(or perhaps $512^3$). For colloidal particles to move a significant distance in this time, their velocities $v_{sed}$ must be of order $10^{-3}$ LU or more. However, the maximum safe viscosity is of order $1$ LU; the colloid radius $a$ is a few LU; and the density $\rho = 1$; therefore Re $ = v_{sed}\rho a/\eta$ must be set at about $10^{-2}$, or else no colloids will move at all far during the course of the simulation. Re can be reduced from this by perhaps one or two orders of magnitude by reducing $v_{sed}$, but only at the cost of very long simulation times. Even this does not approach the tiny values of Re often encountered for real colloids.

However, in most problems, this does not matter: all Reynolds numbers below an appropriate (problem-dependent) threshold are already virtually equivalent. To see this, note that at low enough Re, the velocity field around a sedimenting particle is quasi-static; this Stokesian limit represents diffusive equilibrium of momentum. In principle one could use a variety of algorithms to solve the quasi-static problem. But whatever method was used, it would not be appropriate to do this to accuracy of one part in $10^7$, when there are various percent-level errors arising from other aspects of the code (such as the discretisation of the colloid links in the bounceback algorithm, and in the treatment of interfacial physics). Thus, the goal of LB should not be to simulate `fully realistic' Re
(of order $10^{-7}$) but to use values of Re small enough that the resulting error, in the colloid and solvent dynamics, is no more than a few percent. Such values represent a `realistic' but not a `fully realistic' simulation.

To explore this further,
Figure \ref{benchsphere} shows contour plots for the magnitude of the difference $\Delta {\bf u}$ in the normalised velocity field ${\bf u} = {\bf v}/v_{sed}$ for various simulations of a single falling sphere (with periodic boundary conditions). Our `reference' simulation has very small Re $= 3\times 10^{-6}$. The flow is computed by working in the co-falling frame so that the velocity field converges to a time-independent limit on the lattice; this can be calculated with high accuracy. (Note that such a frame cannot be defined for a problem involving {\em more} than one sphere.) The resulting contour lines for ${\bf u}$ itself are shown in the first panel; difference plots for Re = 0.008, 0.08 and 0.8 are then presented. The latter simulations did not use the co-falling frame; the two higher values are in a range realisable for multi-sphere problems, using reasonable run times. We find that for Re = 0.08, $|\Delta {\bf u}| \le 0.03$ through the main bulk of the fluid, with a slightly larger error close to the colloid. It is possible that this error is partly from the sudden shape changes of the particle as it crosses the lattice (rather than directly from the finite diffusivity of the momentum field). In any case, the deviation from the zero Reynolds number mobility for the falling sphere is about 2\%. This suggests that Re $< 0.1$ represents an acceptably small value for most colloid simulations, and that somewhat larger values might even be acceptable (at least for qualitative exploration of parameter space prior to large-scale production runs; even at Re = 0.8 we find an error in $v_{sed}$ of only about 5\%). This concurs with the comments of Batchelor \cite{batchelor} that in flow past a sphere, all Re $< 1$ are practically equivalent. But note that Batchelor's remark addresses the nonlinear term in Navier Stokes only, whereas our
error stems also from the time derivative, since our flow is nonstationary in the lattice frame. 

In conclusion, for most problems of interest, the level of systematic error caused by Re $< 0.1$, or more conservatively Re $\le 0.05$, is probably acceptable. One possible exception lies in the study of velocity fluctuations in steady sedimentation at large length scales \cite{segre}, whose subtle physics could amplify systematic errors that are negligible at the scale of a single colloid. (For lattice Boltzmann work on this topic see \cite{laddsediment}.)

\begin{figure}
\begin{center}
\includegraphics[width=12cm]{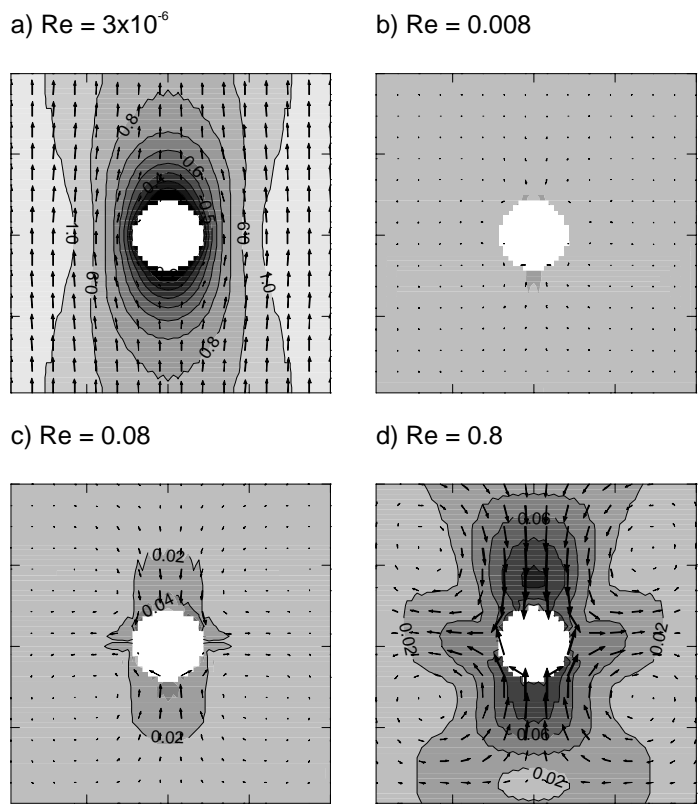}
\end{center}
\caption{\label{benchsphere} Contour plots showing magnitude of the normalized velocity field ${\bf u}$ for a reference simulation at very low Re = $3\times 10^{-6}$ (top left); and velocity difference fields $|\Delta{\bf u}|$ for
Re = 0.008, 0.08, and 0.8. These are for a sphere undergoing sedimentation in periodic boundary conditions. Reference case: contour interval 0.1. Other plots: contour interval 0.02. The arrows denote the velocity field / velocity difference field in each case (scaled similarly). } 
\end{figure}

We turn now to the role of the gravitational capillary number, Ca. We showed in Figure \ref{bench} a series of snapshots for colloidal sedimentation in binary fluids with a stratified interface, at Ca = 280 and Re = 0.02. The simulated Ca is large, but still much smaller than would be typical for micron-scale colloids. Indeed, for typical interfacial tension $\sigma$ between two fluids, to achieve this small a Ca in ordinary gravity
would involve using millimetre sized particles -- so heavy that the appropriate Re is actually {\em more} than the simulated one, unless the solvent viscosity is of order 1000 times larger than that of water. (Large particles in viscous solvents are studied in \cite{segre}, and certainly do show interesting properties.) For viscosities similar to water, our parameters correspond to
micron-scale particles, but in an ultra-high gravitational field (of order $10^6$ times the usual one, if maximal density mismatch is used). 

However, this insistence on `fully realistic' matching of dimensionless parameters between the simulations and laboratory is misguided. Judging from the sedimentation test reported above, Re = 0.02 is small enough to represent any small value; numerical effort is only wasted by reducing it further. Similar remarks probably apply to Ca also. Thus, in principle we should increase $\sigma$ by many orders of magnitude if we want to place Ca within the realistic experimental range for micron-scale colloids. But this would involve taking $\sigma$ far beyond any values benchmarked in \cite{kendon}, and success appears unlikely, since the hugely enhanced interfacial forces would probably lead to numerical instability. However, it may well be that Ca = 280 is already large enough to reproduce, at least for our chosen benchmark problem, all the realistic colloid physics of the large Ca limit. Perhaps, in fact, we can afford to reduce Ca by one or two orders of magnitude and still see the same physics. (Indeed, a similar run with Ca around 20 and Re = 0.2 gives no great differences from Figure \ref{bench}; data not shown.) Similar benchmarking issue will be important to a number of other problems involving colloids in binary solvents, and we leave it open to future study to find out how small a value of Ca is acceptably large, in this and related cases.

In summary, we can state that LB has a good prospect of achieving `realistic' simulations in the sense of a proper relative hierarchy of importance among competing physical effects (capillary terms, diffusion, and inertia, as governed respectively by Ca, 1/Pe, Re). But there is very little prospect of achieving `fully realistic' simulations in the sense of actually resolving the several decades in time and length, between one effect and the next, that are present in typical laboratory experiments. The reason for this is primarily to do with computational resources, and it affects competing algorithms as well as LB. Currently then, no simulation method for colloid hydrodynamics problems can be used reliably, unless it is supported by some physical insight into the nature of the problem being solved. 

\section{Adding noise to LB}
\label{ghosts}
There are many physics problems involving colloid hydrodynamics where Brownian motion is important. The latter arises from the bombardment of the colloids by random forces from the surrounding fluid, and in the well-known simulation method of Brownian dynamics one simply adds a random force directly to each colloid \cite{brownian}. However, this neglects many-body hydrodynamics which, among other things, induces nontrivial correlations between the noise forces acting on different colloidal particles. The proper way to deal with noise in LB is to add random forces to the fluid itself and allow these to propagate, via the hydrodynamic fields, into the colloid sector. 

At the continuum level, this amounts to adding a fluctuating stress
to Eq.\ref{CONSTITUTIVE} \cite{landau-sp1}:
\begin{equation}
\label{CONSTITUTIVENOISE}
\Pi_{\alpha\beta}=g_{\alpha}v_{\beta}+p\delta_{\alpha\beta}-\eta_{\alpha\beta\gamma\delta}\nabla_{\gamma}v_{\delta}+s_{\alpha\beta}
\end{equation}
The fluctuating stress 
$s_{\alpha\beta}$ is a zero-mean Gaussian
random variable whose variance, for a fluid at temperature $T$, is fixed by the
fluctuation-dissipation theorem (FDT) to be $\langle s_{\alpha\beta}({\bf
x},t)s_{\gamma\delta}({\bf
x'},t')\rangle=2k_BT\eta_{\alpha\beta\gamma\delta}\delta({\bf x-x'})\delta(t-t')$.
One way forward, due to Ladd \cite{ladd}, then consists of adding a corresponding stochastic term to the microscopic stress tensor which enters the equilibrium distribution
in Eq.\ref{LBE}. However, the numerical results from this are not accurate. 
This is because of the non-hydrodynamic degrees of
freedom that (in order to maintain Galilean invariance and isotropy) are necessarily retained within the LB method alongside the hydrodynamic ones: the `ghosts'. 
If the noise terms act only on the hydrodynamic modes, the ghosts continually
drain thermal energy away so that the hydrodynamics never reaches equilibrium.

An improved method is presented in \cite{adhikari}, in which we promote the LBE, Eq.\ref{LBE}, into a discrete Langevin equation where the $f_i$ are interpreted as instantaneous, fluctuating densities in phase space:
\begin{equation} \label{FLBE}
f_i({\bf x}+{\bf c}_i,t+1)=f_i({\bf x},t)+{L}_{ij}(f_i({\bf x},t)-f_i^{0}({\bf x},t))+\xi_i({\bf x},t)
\end{equation}
with noise terms $\xi_i({\bf x},t)$. To recover thermal equilibrium, the $\xi_i$
{\em must} be linked, by a fluctuation-dissipation theorem (FDT), to the collisional dissipation. 
The derivation of the required FDT is quite subtle \cite{adhikari} and
in practice requires a careful analysis of all sources of dissipation within the 
collision process, through a study of its eigenmodes. There are nontrivial correlations among the $\xi_i$ at a given site and timestep: specifically these must be interdependent in such a way as to exactly conserve $\rho$ and $ g_{\alpha}$.  For a general LB scheme in $d$ dimensions containing $n$ velocities (a `D$d$Q$n$ model'), there are precisely $n$ eigenvectors, corresponding to the $n$ degrees of freedom contained in the $f_i$ at a given site.  
A complete mode count then consists of one null eigenvector corresponding to the
conserved density $\rho$; $d$ null eigenvectors corresponding to the $d$ conserved components of the momentum $g_{\alpha}$;
$\frac{1}{2}d(d+1)$ eigenvectors
corresponding to the deviatoric momentum flux; and the remaining
$n-(1+d+\frac{1}{2}d(d+1))$ ghost mode eigenvectors. 

We can formally set
$\xi_i = \xi_i^H + \xi_i^G$, with $H$ the hydrodynamic subspace and $G$ its
complement, the ghost subspace. Here $\xi_i^H$ produces thermal fluctuations in the stress tensor, and is the noise used by Ladd \cite{ladd}.  The remaining terms are $\xi_i^G$: these maintain thermal equilibrium for the ghosts.
An explicit method for constructing appropriate noise is described in \cite{adhikari}. In practical simulations, we continue to set mass, length, and time units
so that $\rho=1$ on an unit lattice, and $c_s^2=k_BT/\mu={1/3}$, where $\mu$
is the mass of one of our fictitious fluid particles. We then choose $\rho/\mu={\cal N} \gg 1$ as the mean
number of particles per lattice site. (If this inequality is
not satisfied, the fluid ceases to be a continuum at the lattice scale.)
Since the LB fluid is in fact an ideal gas, fluctuations then obey $\langle\delta{\cal N}^2\rangle={\cal N} = c_s^2\rho/k_BT$.  

The consistency of our approach can be assessed by measuring numerically the
`equilibration ratio' for fluctuating hydrodynamic quantities. This is the
ratio of an actual variance (of, say, a Fourier amplitude of momentum) to the
one required by thermal equilibrium at the temperature $T$ chosen for
the simulation. As shown in \cite{adhikari} our method gives equilibration ratios within a few percent of unity, whereas omitting $\xi_i^G$ gives discrepancies of order 30\%. Note that in the method of \cite{dunweg}, the dissipative coupling between colloids (treated as pointlike) and fluid is accompanied by noise terms which may swamp those arising directly from the fluid itself. This could resolve the problem with equilibration -- or it could merely mask it. In particular, correlations between different colloidal particles must depend on the proper transfer of noise forces through the solvent; so one should carefully check for errors in these quantities.

Fig.\ref{colloid} shows the impulse response function $r(t) = \langle
v(t)\rangle/ v(0)$, and velocity autocorrelator $c(t) = M\langle v(0)v(t) \rangle/kT$
for a Cartesian component $v(t)$ of the velocity of a colloidal particle of
mass $M$. The colloid is suspended in a quiescent fluctuating fluid of equal
density, in three dimensions on a D3Q15 lattice \cite{adhikari}. This is a
parameter-free comparison; FDT demands $r(t) = c(t)$. We find excellent
agreement for time intervals beyond a few timesteps; in fact $c(t+\Delta) = r(t)$
to high accuracy even at short times (not shown), where the best offset $\Delta$ depends slightly on parameters, but is about $0.5$. We attribute the offset to imperfect
resolution of a rapid sound-mediated decay at very short times. (This is
singular for an incompressible fluid, with $c(0+) = 2/3$.) The offset causes a
slight deficit, at most a few percent, in the colloidal self diffusion constant $D=\int_0^\infty
\langle v(0)v(t)\rangle dt$ from its FDT value. There is a
bigger deficit in the equal time correlator $c(0)$, but even this becomes
accurate for larger colloids, where the fast decay is better resolved. (For practical colloid simulation, $D$ matters rather than
$c(0)$.) The inset of Fig. \ref{colloid} shows corresponding data for rotational degrees of freedom; agreement is again excellent. We conclude that, although
some issues remain concerning the choice of optimal parameters and run times for
problems involving Brownian motion of colloids, the fluctuating LBE presented in \cite{adhikari} represents a very promising way forward for simulating such problems.

\begin{figure}
\begin{center}
\includegraphics[width=10cm]{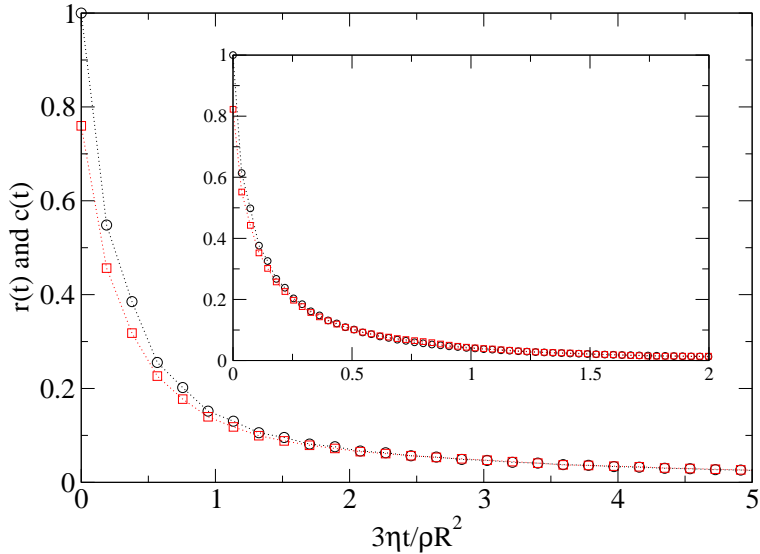}
\end{center}
\caption{\label{colloid} Velocity response (circles) and correlator (squares)
for a colloid. Colloid radius $R=2.3$; $\eta=1/3, {\cal N}=400$.
D3Q15 lattice (BGK collision matrix). Inset: Angular velocity response (circles) and correlator (squares); $R = 3.71, \eta = 1/6, {\cal N} = 400$. Correlators represent data collected over 20000 timesteps, averaged over 3 cartesian velocity components and 10 colloidal particles.} 
\end{figure}

\section{Conclusion}
In this paper we have outlined recent progress in the study of colloid hydrodynamics using lattice Boltzmann. We have made significant progress in the description of colloids in binary solvents and also for colloids in thermal solvents undergoing Brownian motion. Both of these aspects of the code appear to behave sensibly, and in the near future we hope to perfect these tools and apply them to a variety of interesting scientific questions beyond the simple benchmark problem of Figure \ref{bench}. 

Acknowledgements: Work funded by EPSRC GR/R67699 (RealityGrid). MEC thanks B. Dunweg for valuable discussions during the CODEF meeting.


\end{document}